\documentclass[aps,a4paper,superscriptaddress,nofootinbib,twocolumn]{revtex4-1}
\usepackage{amsmath,amssymb,graphicx,multirow,dcolumn,bm,latexsym,soul,nicefrac}
\usepackage[colorlinks,linkcolor=blue,citecolor=blue,urlcolor=blue ]{hyperref}
\usepackage{acronym}
\usepackage{subfigure}
\usepackage{appendix}
\usepackage{longtable}
\usepackage{footnote}
\usepackage{xcolor}
\usepackage{ulem} 
\newcommand{\be}{\begin{equation}}
\newcommand{\ee}{\end{equation}}
\newcommand{\bea}{\begin{eqnarray}}
\newcommand{\eea}{\end{eqnarray}}
\newcommand{\nn}{\nonumber}

\newcommand{\TSS}{MOE Key Laboratory of TianQin Mission, TianQin Research Center for Gravitational Physics
$\&$ School of Physics and Astronomy, Frontiers Science Center for TianQin, Gravitational Wave Research Center of CNSA, 
Sun Yat-sen University (Zhuhai Campus), Zhuhai 519082, China}

\newcommand{\PU}{Astronomy Department, School of Physics, Peking University, Beijing 100871, China}
\newcommand{\Kavli }{Kavli Institute for Astronomy and Astrophysics, Peking University, Beijing 100871, China}
\newcommand{\CU}{Institute of Astronomy, University of Cambridge, Madingley Road, Cambridge CB3 0HA, UK}
\def\apj{Astrophysical Journal}  
\def\apjl{Astrophysical Journal Letters} 

\def\mnras{MNRAS}                
\def\prd{Phys.~Rev.~D}           
\def\prl{Phys.~Rev.~Lett}        
\def\araa{ARA\&A}                
\def\nat{Nature}                 
\def\aap{A\&A}                   
\def\aapr{A\&A~Rev.~}            
\def\pasp{PASP}                  

\def\jcap{JCAP}

\allowdisplaybreaks

\begin{document}
\title{Measuring the Hubble Constant Using Strongly Lensed Gravitational Wave Signals}

\author{Shun-Jia Huang}
\affiliation{\TSS}

\author{Yi-Ming Hu}
\email{corresponding author: huyiming@sysu.edu.cn}
\affiliation{\TSS}

\author{Xian Chen}
\affiliation{\PU}
\affiliation{\Kavli}

\author{Jian-dong Zhang}
\affiliation{\TSS}

\author{En-Kun Li}
\affiliation{\TSS}

\author{Zucheng Gao}
\affiliation{\CU}

\author{Xin-yi Lin}
\affiliation{\TSS}

\date{\today}

\begin{abstract}
The measurement of the Hubble constant $H_0$ plays an important role in the study of cosmology. 
In this work, we propose a new method to constrain the Hubble constant using the strongly lensed gravitational wave (SLGW) signals. 
Through reparameterization, we find that the lensed waveform is sensitive to the $H_0$.
Assuming the scenario that the electromagnetic counterpart of the GW source can be identified, our method can give meaningful constraints on the $H_0$. 
We then apply Fisher information matrix and Markov Chain Monte Carlo to evaluate the potential of this method.
For the space-based GW detector, TianQin, the $H_0$ can be constrained within a relative error of $\sim$ 1\% with a single SLGW event.
\end{abstract}

\maketitle
\acrodef{SLGW}{strongly lensed gravitational wave}
\acrodef{GW}{gravitational wave}
\acrodef{LVK}{LIGO-Virgo-KAGRA}
\acrodef{EM}{electromagnetic}
\acrodef{GL}{gravitational lensing}
\acrodef{SNe Ia}{type-Ia supernovae}
\acrodef{CMB}{cosmic microwave background}
\acrodef{BBH}{binary black hole}
\acrodef{MBBH}{massive binary black hole}
\acrodef{BNS}{binary neutron star}
\acrodef{EMRI}{extreme mass ratio inspiral}
\acrodef{GDWD}{Galactic double white dwarf binary}
\acrodef{AGN}{active galactic nuclei}
\acrodef{SNR}{signal-to-noise ratio}
\acrodef{FIM}{Fisher information matrix}
\acrodef{MCMC}{Markov Chain Monte Carlo}
\acrodef{PSD}{power spectral density}
\acrodef{GR}{general relativity}
\acrodef{PN}{post-Newtonian}
\acrodef{SPA}{stationary phase approximation}
\acrodef{PM}{point mass}
\acrodef{SIS}{singular isothermal sphere}

\section{Introduction}\label{sec:intro}


The Hubble constant has been independently measured by various methods with ever-increasing measurement precision in recent years. 
However, \ac{SNe Ia} observations consistently prefer a higher value of $H_0 = 73.04\pm1.04 \mathrm{\ km\ s^{-1}\ Mpc^{-1}}$ \cite{Riess:2022}, while the \ac{CMB} observations prefer a lower value of $H_0 = 67.4\pm0.5  \mathrm{\ km\ s^{-1}\ Mpc^{-1}}$ under the flat $\Lambda$ cold dark matter ($\Lambda$CDM) model \cite{Planck:2020}.
The discrepancy between these two reference measurements is thus established at 5$\sigma$, which is recognized as the ``Hubble crisis" \cite{Valentino:2021,Riess:2022}. 


Ever since the first direct detection in September 2015, 90 \ac{GW} events have been reported by the \ac{LVK} collaboration \cite{Abbott:2019,Abbott:2021,Abbott:2021b,Abbott:2021c,Abbott:2021d}, which made the \ac{GW} observation a new window to understanding the Universe.
\ac{GW} signals provide direct measurements of the luminosity distance, and combined with the redshift information, they can serve as standard sirens to map the evolution of the Universe \cite{Schutz:1986,Holz:2005,Nissanke:2013,Abbott:2017b}.
The latest LVK measurement yields $H_0 = 68^{+8}_{-6} \mathrm{\ km\ s^{-1}\ Mpc^{-1}}$ using 47 GW events from the third \ac{LVK} Gravitational-Wave Transient Catalog (GWTC–3) \cite{Abbott:2021e}.


If the \ac{GW} from a coalescing binary passes near massive objects, \ac{GL} will affect the \ac{GW} in the same way as it does for light \cite{Wang:1996,Nakamura:1998,Takahashi:2003}, 
which will influence the strain of \ac{GW}, and, in strong lensing cases, produce multiple images with arrival time delay.
It has been proposed that the \ac{GL} can also be used to study cosmology by measuring the difference in arrival time from multiple images and some lens properties \cite{Refsdal:1964,Treu:2010}. 
This idea has been applied in the \ac{EM} domain to infer the value of $H_0 = 73.3^{+1.7}_{-1.8} \mathrm{\ km\ s^{-1}\ Mpc^{-1}}$ from a joint analysis of six lensed quasars in flat $\Lambda$CDM \cite{Treu:2016,Wong:2020}.
Several searches for \ac{GW} lensing signatures have already been performed during the first three observing (from O1 to O3) runs (see \cite{Hannuksela:2019,LIGO:2021,Diego:2021} and the references therein), and there is still no officially confirmed lensing signal by \ac{LVK}.
The lensed \ac{GW} signals are also expected to be detected by space-based gravitational wave detectors such as LISA \cite{Amaro:2017,Gao:2022}.


Recently it's been realized that time delays between multiple signals of lensed \acp{GW} by the future \ac{GW} detections can be used to measure the Hubble constant \cite{Sereno:2011,Liao:2017,Wei:2017,Li:2019,Cremonese:2020,Hannuksela:2020,Cao:2021,Hou:2021,Qi:2022}. 
Liao et al.~\cite{Liao:2017} reported a waveform-independent strategy by combining the accurately measured time delays from \ac{SLGW} signals with the images and lens properties observed in the \ac{EM} domain and shows that 10 such systems are sufficient to constrain the Hubble constant within an uncertainty of $0.68\%$ for a flat $\Lambda$CDM universe in the era of third-generation ground-based detectors.

In this work, we propose a new method to measure the Hubble constant $H_0$ using \ac{SLGW} signals.
The waveform is reparameterized to explicitly include $H_0$ in the parameter set.
We find that the \ac{SLGW} waveforms are sensitive to $H_0$.
This was made possible by utilizing the waveform \emph{per se}, instead of just the time delay or the magnification \cite{Sereno:2011,Liao:2017,Wei:2017,Li:2019,Cremonese:2020,Hannuksela:2020,Cao:2021,Hou:2021,Qi:2022}.
As a result, our method is capable of measuring $H_0$.

We evaluate the potential of this new method in measuring the Hubble constant and provide preliminary results for TianQin, which is a planned space-based \ac{GW} observatory sensitive to the millihertz band \cite{Luo:2016}.
In recent years, a significant amount of effort has been put into the study and consolidation of the science cases for TianQin \cite{Hu:2017,Wang:2019,Feng:2019,Bao:2019,Shi:2019,Liu:2019,Fan:2020,Huang:2020,Mei:2021,Zi:2021,Liang:2021,Liu:2022,Zhu:2022a,Zhu:2022b,Zhang:2022,Lu:2022,Sun:2022,Xie:2022}. 
For the detected \ac{GW} sources, TianQin's sky localization precision can reach the level of 1 deg$^2$ to 0.1 deg$^2$ \cite{Wang:2019,Liu:2019,Fan:2020,Huang:2020}, which makes it possible to combine subsequent \ac{EM} observations to implement multi-messenger astronomy.
For an order-of-magnitude estimation of the \ac{SLGW} detection rate, we adopt the detecting probability of \ac{SLGW} from \ac{MBBH} mergers as $\sim 1\%$ for space-based gravitational wave detector \cite{Gao:2022}, while under the optimistic model, the detection rate of \ac{GW} from \ac{MBBH} is $\gtrsim O(10^2)/yr$ \cite{Wang:2019}. 
Therefore the total detection number for \ac{SLGW} events can be as high as $\gtrsim O(5)$ over the five-year mission lifetime.
We consider two cases: a) \ac{GW} source only, and b) \ac{GW} source and lens simultaneously.
These cases represent different levels of information available, we assume for both cases that the \ac{EM} counterpart of \ac{GW} source is observed.
This can be justified as some studies have argued that if the massive binary black hole evolve in gas-rich environment, the accretion of gas could produce \ac{EM} radiation \cite{d'Ascoli:2018}, which can be used to determine the source redshift \cite{Tamanini:2016}.
Our calculations show that the Hubble constant can be well constrained using our method in these two cases. 

This manuscript is organized as follows. 
In section \ref{sec:theory}, we describe our method to measure $H_0$ using \ac{SLGW} signals.
In section \ref{sec:method}, we introduce the parameter estimation method used in this paper.
In section \ref{sec:result}, we evaluate the expected measurement precision of $H_0$ using our new method.
Finally, we summarize and discuss in section \ref{sec:sum}.
Throughout this manuscript, we use $G = c = 1$ and assume a flat $\Lambda$CDM cosmology with the parameters 
$\Omega_M = 0.315$, $\Omega_{\Lambda} = 0.685$.
For the simulation, we also adopt $H_0 = 67.66 \mathrm{\ km\ s^{-1}\ Mpc^{-1}}$ \cite{Planck:2020}.

\section{The \ac{SLGW} signals}\label{sec:theory}

The strong \ac{GL} not only alters the \ac{GW} amplitude and phase but also produces multiple duplicates of the same signal. 
The calculation of the \ac{SLGW} waveform considering the detector response is complicated and generally speaking, there is no compact form.
However, analytical description is available when certain simplifications are adopted.
For example, in the case of the \ac{PM} lens model, and utilizing the geometric optics approximation (which means that the Schwarzschild radius of the lens is much larger than the wavelength of the \acp{GW}), one can express the waveform as a superposition of two \ac{GW} signals. 
This can be derived by applying the stationary phase approximation, which is valid as the \acp{GW} phases evolve much faster than their amplitude.
One can derive \cite{Takahashi:2003} 
\bea
{\widetilde{h}}_\alpha^L\left(f\right)  &=&  \bigg[ |\mu_+|^{1/2} \Lambda_{\alpha}(t) e^{ -i(\phi_D + \phi_{p,\alpha})(t) } \nn\\
&& -i|\mu_-|^{1/2} e^{2\pi i f t_{d}} \Lambda_{\alpha}(t+t_{d})  e^{ -i(\phi_D + \phi_{p,\alpha})(t+t_{d}) } \bigg]  \nn\\
&&   \times \mathcal{A} f^{-7/6} e^{i\Psi(f)} , \qquad\qquad\qquad  \alpha=I,II 
\label{eq:lensed waveform}
\eea
where $\alpha=I,II $ denotes the index of two interferometer. 
In this proof-of-principle study, for the convenience and efficiency of calculation, we use this analytical waveform derived from the point-mass model and geometric optics approximation to simulate the \ac{SLGW} signals.
The GW amplitude and phase are 
\bea
\mathcal{A}= \sqrt{\frac{5}{96}} \frac{\pi^{-2/3}\mathcal{M}_z^{5/6}}{D_S(1+z_S)^2}
\label{eq:amp}
\eea
and 
\bea
\Psi(f) = 2 \pi f t_\mathrm{c} - \phi_\mathrm{c} - \frac{\pi}{4} + \psi_{\mathrm{PN}},
\label{}
\eea
where $D_S$ is the angular diameter distances from the source to the observer, $D_S(1+z_S)^2$ is the luminosity distance to the source\footnote{In order to avoid possible confusion, we express all distances in terms of angular diameter distance.}, and $\psi_{\mathrm{PN}}$ is the \ac{PN} phase as a function of the redshifted chirp mass $\mathcal{M}_z=(1+z_S)(M_1 M_2)^{3/5}/(M_1+M_2)^{1/5}$ and symmetric mass ratio $\eta=(M_1 M_2)/(M_1+M_2)^2$.
In this work, we expand the \ac{PN} phase to 2 PN order (see \cite{Feng:2019} and the references therein).

The initial frequency of the waveform is
\bea
f_{\mathrm{in}}=(5/256)^{3/8} \frac{1}{\pi} \mathcal{M}_z^{-5/8} (t_c-t)^{-3/8},
\label{}
\eea
and we adopt the cutoff frequency at the innermost stable circular orbit (ISCO)
\bea
f_{\mathrm{ISCO}}= (6^{3/2}\pi M_z)^{-1},
\label{}
\eea
where $M_z=(1+z_S)(M_1+M_2)$ is the redshifted total mass. 

The detector response, the polarization phase, and the Doppler phase can be defined by
\bea
\Lambda_\alpha(t)  &=&  \sqrt{ (1+\cos^2\iota )^2{F_\alpha^+(t)}^2+4\cos^2\iota {F_\alpha^\times(t)}^2}, \\
\phi_{P,\alpha}(t)     &=& {\tan}^{-1}\ \left[\frac{{2\cos\iota\ F}_\alpha^\times(t)}{{(1+\cos^2\iota)F}_\alpha^+(t)}\right],
\eea
and
\bea
\phi_D (t) = 2\pi fR\sin\left(\frac{\pi}{2}-\beta\right)\ \cos(2\pi f_\mathrm{m}t-\lambda),
\eea
with $R = 1$ AU and $f_m = 1/\mathrm{year}$. 
$F_\alpha^{+, \times}(t)$ is the antenna pattern functions of the source position ($\beta, \lambda$) and polarization angle $\psi_S$ \cite{Huang:2020}, and related to the motion and pointing of the detector.

The magnification factor of each image and the time delay between the double images for the case of the \ac{PM} lens are given by \cite{Takahashi:2003}
\bea
\mu_{\pm}=\frac{1}{2} \pm \frac{y^2+2}{2 y \sqrt{y^2+4}}
\label{eq:magnification factor}
\eea
and 
\bea
t_d&=&4M_L (1+z_L)\left[ \frac{y \sqrt{y^2+4} }{2}+ \ln \left(\frac{\sqrt{y^2+4}+y}{\sqrt{y^2+4}-y}\right)\right], \nn \\
&&
\eea
where $y=D_L\eta_L/(D_S\xi_0)$ is the dimensionless source position in the lens plane, $D_L$ is the angular diameter distances from the lens to the observer, $\eta_L$ and $\xi_0$ are the \ac{GW} source position in the lens plane and the normalization constant of the length, respectively.
When $y\rightarrow \infty$ ($\mu_+=1$ and $\mu_-=0$), Eq. (\ref{eq:lensed waveform}) reduces to the unlensed waveform.

For the case of the \ac{PM} lens, the normalization constant can be given by $\xi_0=\sqrt{4M_LD_LD_{LS}/D_S}$, where $D_{LS}$ is the angular diameter distances from the source to the lens, and $M_L$ is the lens mass.
In this case, the dimensionless source position can be written as
\bea
y= \frac{\eta_L}{2 \sqrt{M_L}} \sqrt{\frac{D_L}{D_SD_{LS}}},
\label{eq:}
\eea
using the distance-redshift relationship \cite{Hogg:1999}
\bea
D= \frac{1}{H_0(1+z)}\int_0^z dz^\prime \frac{1}{E(z^\prime)},
\label{eq:distance-redshift}
\eea
where $E(z^\prime)$ depicts the background evolution of the Universe, it can be found that $y$ is proportional to $\sqrt{H_0}$ in the \ac{PM} lens model. 

For the \ac{SIS} lens model, there are
\bea
\mu_{\pm}= \pm 1+\frac{1}{y}
\eea
and
\bea
t_d&=&8M_L (1+z_L)y,
\eea
where the lens mass inside the Einstein radius is given by $M_L = 4\pi^2v_d^4 D_LD_{LS}/D_S$, and $v_d$ is the velocity dispersion of lens galaxy. 
In this case, the dimensionless source position can be written as
\bea
y= \frac{\eta_L}{4 \pi v_d^2 D_{LS}}.
\label{eq:}
\eea
Using the distance-redshift relationship again, $y$ is proportional to $H_0$ in the \ac{SIS} lens model.



Assuming a flat $\Lambda$CDM model, we can describe a lensed waveform with a set of 13 parameters: the Hubble constant $H_0$, lens mass $M_L$, lens redshift $z_L$, \ac{GW} source position in the lens plane $\eta_L$, \ac{GW} source redshift $z_S$, redshifted chirp mass $\mathcal{M}_z$, symmetric mass ratio $\eta$, inclination angle $\iota$, coalescence time $t_c$, coalescence phase $\phi_c$, \ac{GW} source ecliptic coordinates $\beta$, $\lambda$, and polarization angle $\psi_S$.

\section{Parameter Estimation Methods}\label{sec:method}


To evaluate the potential of measuring the Hubble constant using the \ac{SLGW} signals, we apply both analytical and numerical methods with \ac{FIM} and \ac{MCMC}, respectively.


The \ac{FIM} $\Gamma_{ij}$ is defined as 
\be \Gamma_{ij} = \left(\frac{\partial h}{\partial \theta_i}\bigg|\frac{\partial h}{\partial \theta_j}\right),
\label{eq:FIM} \ee
where $\theta_{i}$ stands for the $i$-th \ac{GW} parameter, 
In the limit that a signal is associated with a large \ac{SNR} $\rho = \sqrt{(h|h)} \gg 1$, 
one can approximate the variance-covariance matrix with the inverse of the \ac{FIM}  $\Sigma = \Gamma^{-1}$, with $\Sigma_{ii}$ describing the marginalized variances of the $i$-th parameter.


Under the framework of Bayesian inference, one would like to constrain parameters $\theta$ with data $d$, or to obtain the posterior distribution $p(\theta|d,H)$ under model $H$.
Under the Bayes' theorem,
\bea
p(\theta|d,H)=\frac{p(\theta|H)p(d|\theta,H)}{p(d|H)} ,
\label{eq:Bayes}
\eea
the posterior is the product of the prior $p(\theta|H)$ and the likelihood $p(d|\theta,H)$, normalized by the evidence $p(d|H)$.
The likelihood of the \ac{GW} signal can be written as 
\bea
p(d|\theta,H) \propto \exp\left[- \frac{1}{2}\bigg(d-h(\theta)\bigg|d-h(\theta)\bigg)\right],
\label{eq:likelihood}
\eea
where the inner product $(\cdot|\cdot)$ defined as \cite{Finn:1992,Cutler:1994}
\bea 
(a|b)=4\Re e\int^\infty_0 \mathrm{d}f \frac{\tilde{a}^*(f)\tilde{b}(f)}{{S}_n(f)},
\label{eq:inner product} 
\eea
and $S_n(f)$ is the one-sided power spectral density, we adopt the formula from \cite{Huang:2020} for the expression of TianQin.
In our case we consider the \ac{SLGW} waveform $h(\theta)$ described by Eq. (\ref{eq:lensed waveform}).
We adopt the emcee \cite{Mackey:2013} implementation of the affine invariant \ac{MCMC} sampler.

\section{Measurement precision of $H_0$}\label{sec:result}

\begin{figure}
		\includegraphics[width=0.5\textwidth]{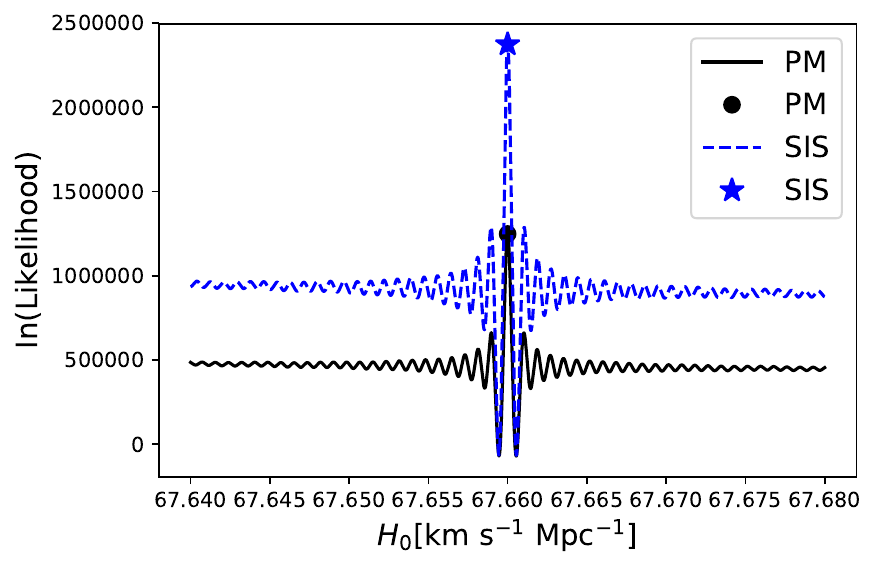}
		\caption{The conditional probability of $H_0$ for \ac{PM} (black solid line) and \ac{SIS} (blue dashed line) lens model, respectively, the black point and blue star indicates the injected value.}
		\label{fig:conditional probability}
\end{figure}


\begin{figure}
		\includegraphics[width=0.5\textwidth]{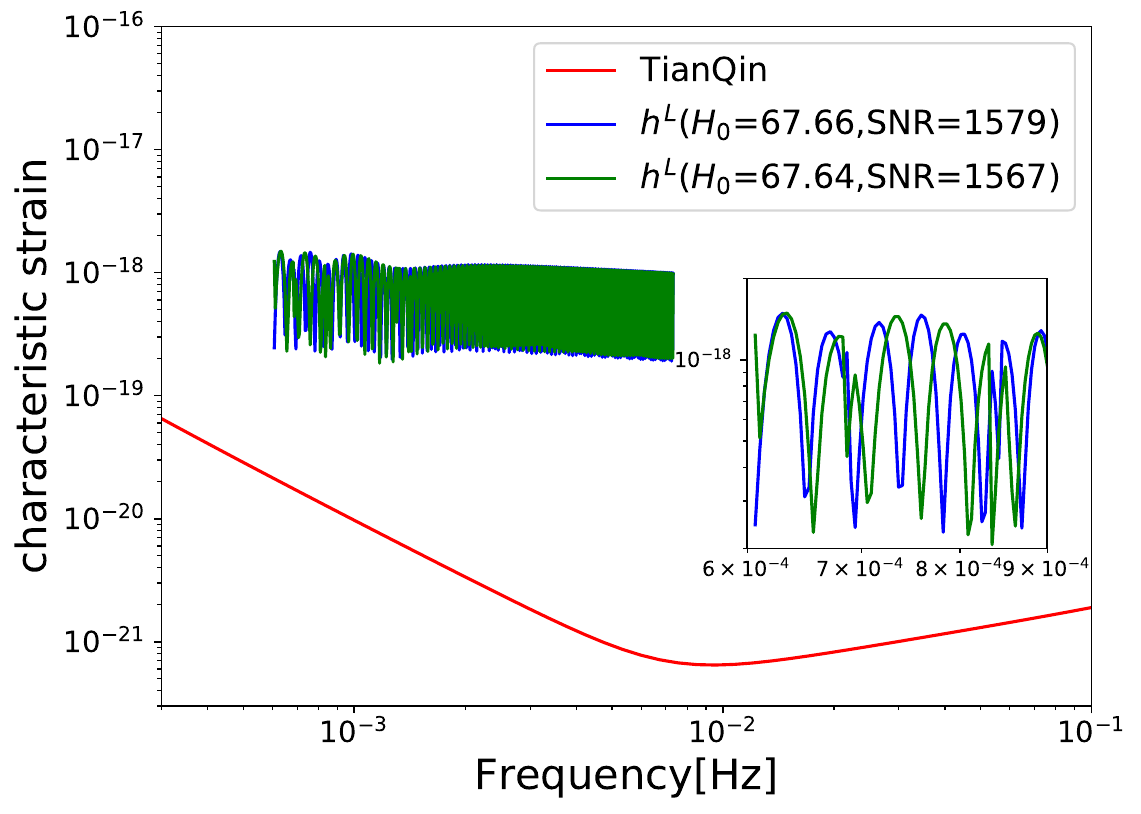}
		\caption{Two strongly lensed waveforms, with all other parameters fixed but $H_0$ slightly different. The values of parameters are listed at the beginning of Sec. \ref{sec:result}}
		\label{fig:Waveform}
\end{figure}

\begin{figure}
		\includegraphics[width=0.5\textwidth]{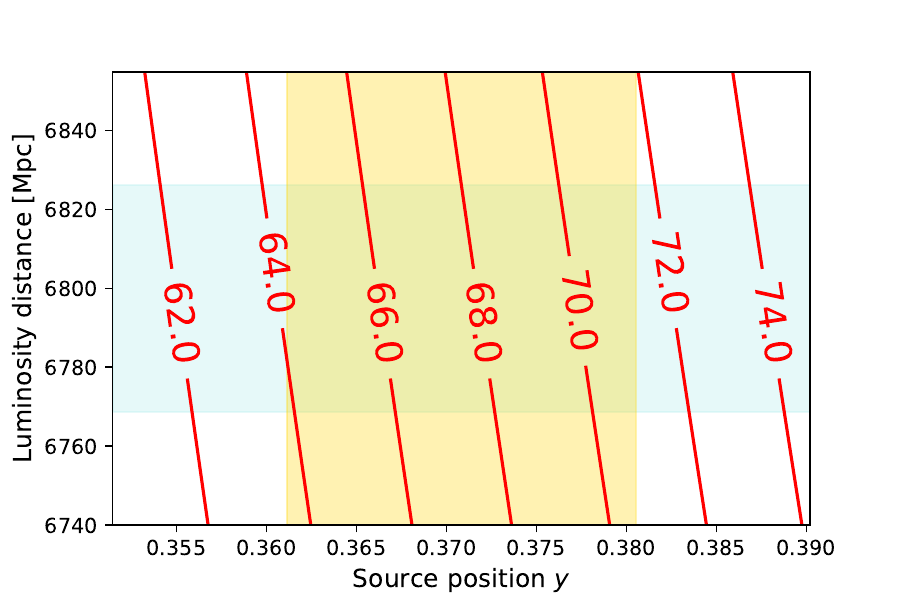}
		\caption{The contour of $H_0$ over different luminosity distances and source positions. The blue and yellow shaded regions represent the measurement error from \ac{FIM} analysis around the true value, showing that measurement of both parameters can be used to constrain the Hubble constant.}
		\label{fig:contour_H0}
\end{figure}

For the lens system parameters, we choose the lens mass $M_L=1\times10^{12} \, \mathrm{M}_\odot$, the lens redshift $z_L=0.5$, the \ac{GW} source position $\eta_L=5$ kpc, and the \ac{GW} source redshift $z_S=1$.
For the remaining \ac{GW} source parameters, we adopt the redshifted chirp mass $\mathcal{M}_z=2.43\times10^{5} \, \mathrm{M}_\odot$, symmetric mass ratio $\eta=0.222$ ($M_1=2\times10^{5} \, \mathrm{M}_\odot$, $M_2=1\times10^{5} \, \mathrm{M}_\odot$), inclination angle $\iota=\pi/3$, coalescence time $t_c=2^{18}$ s ($\sim 3 \ \mathrm{days}$), coalescence phase $\phi_c=\pi/2$, \ac{GW} source ecliptic coordinates $\beta=-4.7^\circ$, $\lambda=120.4^\circ$ (The zenith of TianQin), and polarization angle $\psi_S=\pi/2$.

We first inject a simulated \ac{SLGW} signal $h(H_0^{\rm inject})$ without noise, and match it with a signal with different Hubble constant $h(H_0^{\rm match})$.
The log likelihood is calculated as $\left(h(H_0^{\rm inject})\middle|h(H_0^{\rm match})\right)-\frac{1}{2}\left(h(H_0^{\rm match})\middle|h(H_0^{\rm match})\right)$, In Figure \ref{fig:conditional probability} we demonstrate the sensitivity of log likelihood with respect to the Hubble constant for the \ac{PM} (black solid line) and \ac{SIS} (blue dashed line) lens model, respectively.
It can be found that the sensitivity curves of \ac{PM} and \ac{SIS} models are similar in shape, except that the log likelihood function value of \ac{SIS} model is larger because the magnification of \ac{GL} effect is larger.
For convenience, the \ac{PM} model is used for subsequent calculations.
We observe that the log likelihood varies violently at the level of $0.001 \mathrm{\ km\ s^{-1}\ Mpc^{-1}}$, or $10^{-5}$ for relative uncertainty in Figure \ref{fig:conditional probability}.
The amplitude of the injected lensed waveform as well as a typical template is shown in Figure \ref{fig:Waveform}.
We observe that indeed that the lensed waveform is highly sensitive to the choice of $H_0$ and even a tiny difference in the Hubble constant can lead to significant deviation in the waveform.

Given all other parameters fixed, Figure \ref{fig:conditional probability} and \ref{fig:Waveform} indidcate that the waveform is highly sensitive to the Hubble constant.
However, we do not know other parameters \emph{a priori}, and due to the degeneracy between parameters, if source redshift $z_S$ is not determined, one can not constrain the Hubble constant.
However, studies suggested that space-borne \ac{GW} missions have the potential to accurately localize sources, especially when a network of detectors is considered \cite[e.g.][]{Gong:2021gvw,Liu:2022rvk}.
Therefore, it is fair to assume that the \ac{EM} counterpart might be observe, which will break the degeneracy.
For the following analysis, we choose to fix the other cosmological parameter $\Omega_M$ as we specifically focus on the $H_0$.
We also fix angles like $\beta,\ \lambda$, as any \ac{EM} information about the lens or the host galaxy can help to set these angles \cite{Wong:2020,Dye:2005,Schutz:1986,Bogdanovi:2022}. 
We consider two cases for the different observed EM counterparts: a) \ac{GW} source only, where we fix $z_S$, and b) \ac{GW} source and lens simultaneously, where $z_L$, $M_L$, and $z_S$ are fixed.

In Figure \ref{fig:contour_H0}, we illustrate that by matching template with injected signal, one can constrain both the luminosity distance $D_S(1+z_S)^2$ and the source location $y$, with the associated uncertainties represented by the shaded regions.
From simple derivation one can deduce that by measuring both parameters, it is possible to constrain the Hubble constant.
The \ac{FIM} analysis predicts that the $H_0$ can be constrained to the relative precision to 0.80\% and 0.13\%  for case a) and b), respectively. 

\begin{figure*}
		\includegraphics[width=1\textwidth]{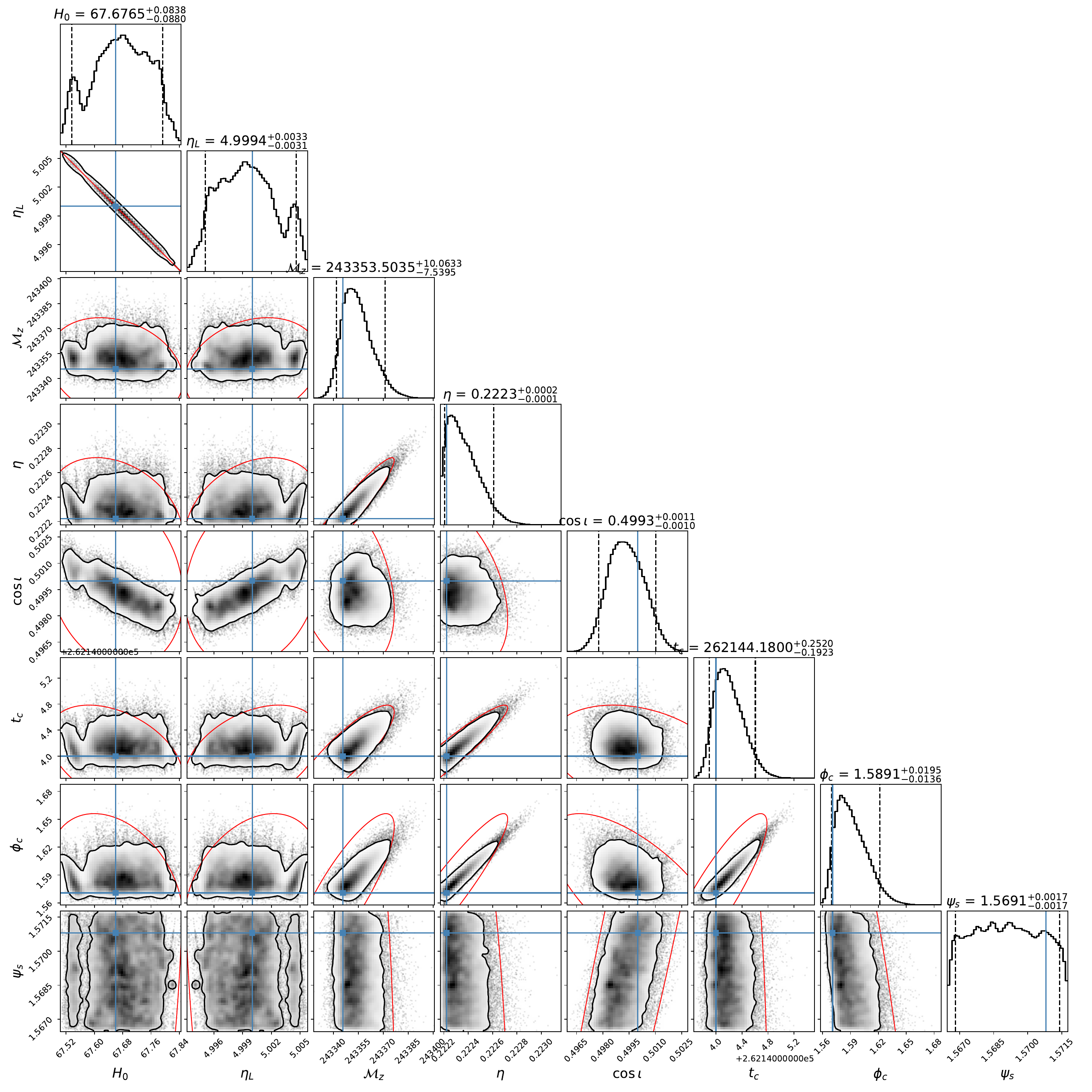}
		\caption{Comparison between the posterior distribution sampled by \ac{MCMC} (black) and the ellipses based on the \ac{FIM} (red) with 90\% credible regions.}
		\label{fig:MCMC}
\end{figure*}

We present an example posterior distribution in Figure \ref{fig:MCMC} from the \ac{MCMC} sampling, with the black contour lines indicating 90\% credible region and we fix source redshift $z_S$, lens redshift $z_L$, and lens mass $M_L$.
We adopt the uniform prior for $H_0 \in (40,100) \mathrm{\ km\ s^{-1}\ Mpc^{-1}}$, $\eta_L \in (0, 10)$ kpc, $\eta \in (0, 0.25)$, $\cos\iota \in (0, 1)$, $\phi_c \in (-\pi, \pi)$, and $\psi_S \in (0, \pi)$.
For $\mathcal{M}_{z}$ and $t_c$, we adopt uniform priors greater than zero.
We check the convergence of the \ac{MCMC} by looking at the Gelman-Rubin statistic \cite{gelman1992inference}, and for all parameters they are consistently smaller than 1.05.

In Figure \ref{fig:MCMC}, we overlap the red ellipses as \ac{FIM} 90\% credible regions for comparison.
It can be observed that the two methods give consistent results, both predict that even with many parameters unfixed, the lensing waveform method can still pinpoint the Hubble constant $H_0$ to the level of 1\%.
Notice that this value is about three orders of magnitude worse than from Figure \ref{fig:conditional probability}. 
This can be explained by the large correlation between $H_0$ and other parameters like source redshift $z_S$ and lens mass $M_L$.
Nevertheless, the 1\% level precision of $H_0$ is still highly encouraging, as it provides an independent measurement from a single event.


\section{Summary and discussion}\label{sec:sum}


In this work, we proposed a new method to measure the Hubble constant using \ac{SLGW} signals.
We reparameterize the waveform to include the Hubble constant in the parameter set.
We find that the space-borne \ac{GW} missions can determine the Hubble parameter to the relative precision of $\sim$1\%. 
Our work extract cosmological parameters from the waveform while many previous works only use information distilled from the waveform, like the time differences between images and magnification factors \cite{Sereno:2011,Liao:2017,Wei:2017,Li:2019,Cremonese:2020,Hannuksela:2020,Cao:2021,Hou:2021,Qi:2022}.
When the \ac{EM} counterparts of source and lens are identified, and the source redshift $z_S$, lens redshift $z_L$, and lens mass $M_L$ are also fixed, one can constrain $H_0$ to the level of $\sim$ 0.1\%. 

We adopted a number of simplifications in this proof-of-principle study.
For example, we only considered the \ac{PM} lens model for lensing effect calculation.
In future we can extend to more general models like the \ac{SIS} or the Navarro-Frenk-White model.
Secondly, we calculate the \ac{GL} effect under the geometric optics approximation \cite{Takahashi:2003}.
For the parameters we choose it is still valid, but a more general treatment would require the consideration of wave optics \cite{Takahashi:2003,Gao:2022}. 
Last but not least, we only considered the \ac{PN} waveform for the inspiral of a quasi-circular, non-spinning compact binary.
Although this hugely saves the computation time, we plan to include a full inspiral-merger-ringdown waveform for a more general binary in the future for more realistic analysis.
We remark though that the discard of merger and ringdown means that a portion of information and \ac{SNR} are not used and we expect a better constraint with the consideration of the IMR waveforms.

\begin{acknowledgments}
We would like to thank Ran Li, Liang-Gui Zhu, Shuai Liu, Xiang-Yu Lv, Jie Gao, Xue-Ting Zhang, Jing Tan for helpful comments and discussions.

This work was supported by the National Natural Science Foundation of China (Grants No. 12173104, 11690022, and 11991053), and Guangdong Major Project of Basic and Applied Basic Research (Grant No. 2019B030302001). 
E. K. L. was supported by the fellowship of China Postdoctoral Science Foundation (Grant No. 2021M703769), and the Natural Science Foundation of Guangdong Province of China (Grant No. 2022A1515011862).
\end{acknowledgments}


\bibliographystyle{apsrev4-1}
\bibliography{Lensing}
\end{document}